\documentclass[fleqn,usenatbib]{mnras}

\usepackage{newtxtext,newtxmath}
\usepackage{subcaption}
\usepackage[T1]{fontenc}
\usepackage{ae,aecompl}

\usepackage{graphicx}	% Including figure files
\usepackage{amsmath}	% Advanced maths commands

\usepackage{hyperref}
\title{New mechanisms for multiple hotspot formation}

\author[M. A. Horton et al.]{
Maya A. Horton,$^{1}$\thanks{E-mail: mh17adw@herts.ac.uk}
Martin G. H. Krause,$^{1}$
and Martin J. Hardcastle$^{1}$
\\
$^{1}$Centre for Astrophysics Research, School of Physics, Astronomy and Mathematics, University of Hertfordshire, College Lane, Hatfield, AL10 9AB, UK\\
}

\date{Accepted XXX. Received YYY; in original form ZZZ}

\pubyear{2023}

\begin{document}
\label{firstpage}
\pagerange{\pageref{firstpage}--\pageref{lastpage}}
\maketitle

\begin{abstract}
    Hotspots of radio galaxies are regions of shock-driven particle acceleration. Multiple hotspots have long been identified as potential indicators of jet movement or precession. Two frequent explanations describe a secondary hotspot as either the location of a prior jet termination point, or a deflected backflow-driven shock: the so-called Dentist's Drill and Splatter Spot models. We created high-resolution simulations of precessing jets with a range of parameters. In addition to the existing mechanisms, our results show three additional mechanisms for multiple hotspot formation: (1) the splitting of a large terminal hotspots into passive and active components; (2) jet stream splitting resulting in two active hotspots; (3) dynamic multiple hotspot complexes that form as a result of jet termination in a turbulent cocoon, linked here to rapid precession. We show that these distinct types of multiple hotspots are difficult to differentiate in synthetic radio maps, particularly hotspot complexes which can easily be mistaken for the jet itself. We discuss the implication for hypothesised binary supermassive black hole systems where jet precession is a key component of the morphology, and show a selection of potential precession candidates found using the LOFAR Two-Metre Sky Survey Data Release 2 (LoTSS DR2). 
\end{abstract}


%%%%%%%%%%%%%%%%%%%%%%%%%%%%%%%%%%%%%%%%%%%%%%%%%%

%%%%%%%%%%%%%%%%% BODY OF PAPER %%%%%%%%%%%%%%%%%%

\section{Introduction}
\label{sec:intro}
Hotspots are considered to be a defining characteristic of Fanaroff \& Riley \citep{fanaroff74} class II (FRII) radio sources \citep{carilli95}. However, from the earliest observations it has been clear that their shape and complexity does not easily fit with simple descriptions of them as planar shocks: whilst there is little doubt that they are regions of shock-driven particle acceleration occurring at the termination points of edge-brightened radio sources \cite[e.g.][]{laing89}, their shape and complexity can be difficult to categorise.

Many FRII sources have multiple hotspots \citep[e.g.][]{laing82,leahy97}, although it is generally expected that a jet can only have one current termination point. Two primary mechanisms for forming multiple hotspots are described in the literature. In the `dentist's drill' model \citep{scheuer82,hardee90,cox91} the jet changes direction, either through precession or through a discontinuous reorientation \citep{ekers78}, on a timescale shorter than that required for the initial hotspot to fade. Under the `splatter-spot' model, a secondary shock is generated by continued supersonic collimated outflow from the initial jet termination point, perhaps after the jet is deflected obliquely by the lobe boundary (the contact discontinuity between jet plasma and shocked ambient medium) \citep{williams85,smith84,lonsdale+barthel86}.

Observationally both of these models have difficulty explaining the diversity of hotspot structures observed in real radio galaxies. Both predict that the jet should currently terminate in a compact `primary' hotspot and that other (`secondary') hotspots should be more diffuse. In the simplest version of the dentist's drill model, the secondary hotspot should show signs of spectral ageing relative to the primary, but this is not observed in some well-studied cases \citep{hardcastle01}; the properties of many secondary hotspots are not consistent with being older, remnant versions of the primaries \citep{valtaoja84}. Models involving a decelerated continued outflow predict that the secondary hotspot should always be downstream of the primary, and should have less efficient particle acceleration, but this is not consistent with observations showing that particle acceleration can be as efficient in the secondary as in the primary \citep{hardcastle07}.  

The formation mechanisms of multiple hotspots are particularly important because of their role as indicators for jet precession or reorientation \citep{krause18}. Jet precession, if caused by the geodetic precession of binary supermassive black holes \citep[e.g.][]{begelman80, gower82} can be used as an indicator of the prevalence of such systems, which are progenitors of high-energy gravitational wave events. Additionally, jet precession can be caused by other mechanisms such as Lense-Thirring or Bardeen-Petterson \citep[e.g.,][]{lense18, bardeen75} and these are also expected to have observational signatures. Previous simulation-based work has often used more or less realistic precession models to generate multiple hotspots \citep[e.g.][]{cox91}.

In a previous paper, \cite{horton20b} used high-resolution hydrodynamic simulations to investigate the effects of precession on the observable properties of jets in radio galaxies, showing that curved, S-shaped and misaligned jets are reliable precession indicators. A systematic study of hotspots was excluded from that paper, given their complexity. Here we report on highest-resolution simulations of such systems that allow us to study the formation of multiple hotspots directly. We identify three novel, precession-driven hydrodynamic processes that may result in the formation of multiple hotspots whose properties would not be inconsistent with observations of spectral ageing and particle acceleration and help to explain discrepancies in size and position between `primary' and `secondary' hotspots. These are in addition to the existing two mechanisms. Finally, we briefly discuss physical mechanisms that can lead to jet precession, and discuss how some of these can have an impact on the search for supermassive binary black hole systems; as \cite{horton20a} highlight, terminal hotspots are particularly important in the use of statistical models to constrain binary separations and gravitational wave strains using precessing jets \citep[see also][]{krause18}.

\section{Methods}
\label{sec:methods}
\subsection{Simulations}
We use the simulation setup described by \cite{horton20b} -- hereafter referred to as H20 -- using PLUTO hydrodynamic code\footnote{\url{http://plutocode.ph.unito.it}} \citep{mignone07} running the HD physics module with a HLLC Riemann solver and second order Runge Kutta (RK2) timestepping with PLUTO's `linear' reconstruction which is second-order accurate in space. Higher-order spatial reconstruction or timestepping significantly increases the computational cost for high-resolution simulations. We chose to work in HD rather than MHD or RHD partly for computational simplicity and the efficient use of a spherical grid: previous work \citep[e.g.][]{hardcastle14} has demonstrated that the use of realistic sub-equipartition fields has no global effect on the lobe or jet dynamics, and although we expect bulk flow speeds to be relativistic in real jets, the flows downstream of shocks that are the topic of this paper will be at most mildly relativistic.

The supersonic ($M=100$) jet was injected as a spot moving on the intersection between a specified precession cone and the inner boundary of the spherical computational volume. The extent of the linear grid in code units was $0.2 \le r \le 5$, $0 \le \theta \le \pi$ and $0 \le \phi \le 2\pi$. This paper uses two of the \_VHR (Very High Resolution) simulation runs described in H20 with a grid setup of 512 x 512 x 1024 in spherical polar coordinates of $r$ (radial), $\theta$ (polar), and $\phi$ (azimuthal) angles respectively. This angular resolution means that the $5^\circ$ injection spots are well resolved with $\sim 300$ grid elements across each spot.

The initial jet density and pressure are set to match that of the ambient environment but the conical jet injection described by H20 has the feature that the jet naturally expands, then recollimates and heats up within a short distance from the injection location, as expected in theoretical models \citep{kaiser97}, while remaining supersonic ($M \sim 10$). The jets and lobes on large scales are therefore hot and low-density with respect to their environment, matching at least qualitatively what is expected for real radio galaxies \citep{hardcastle13}. As discussed by H20, the simulations are intrinsically scale-free but if we assume a plausible environment for radio galaxies and set the outer radius of the simulations to 300 kpc, then the simulation unit of distance is 60 kpc, the sound speed is 480 km s$^{-1}$ and one simulation time unit is $1.2 \times 10^8$ years, while the time step between saved simulation volumes is $1.2 \times 10^5$ years. The jet power is then well matched to the expected powers of FRII radio galaxies.

The parameters varied were precession period $pp$ and precession cone opening angle $\psi$. All jets had a fixed jet half-cone angle $\phi$ of $5^\circ$; $pp$ was set at either 1 or 5 turns per simulation time (i.e. precession periods in physical units, using the conversions above, in the range $10^7$--$10^8$ years) and $\psi$ was set at $45^\circ$. These runs were selected from H20 because the wide precession cone opening angle gave rise to interesting behaviour regardless of how many turns there were over the lifetime of the source. Because of this choice we are modelling relatively slow precession but, as discussed by H20, we expect that the qualitiative behaviour we observe in simulated radio morphology will not be strongly dependent on the precession period. We did not vary the injection Mach number $M$ nor examine intermediate values of $\psi$ as set out in H20 as this would have required additional very high resolution (VHR) runs. As in the previous paper jets were injected into a uniform environment. We did not run jets for more than one simulation time. No other parameters were altered for \_VHR runs (in the bulk of this paper we analyse only 45\_100\_1\_VHR and 45\_100\_02\_VHR, out of the four original simulations). We also ran a very high resolution straight jet, 15\_100\_STR\_VHR, to see whether the same hydrodynamic effects appeared without precession. Movies showing the evolution of the pressure contours for these simulations can be found at \url{https://uhhpc.herts.ac.uk/~mayaahorton/multihotspots.html}.

We recognise that the term `hotspot' is often closely associated with flat spectral indices; the spectral index of the radio galaxies are not modelled here. However, modelled emissivity is a strong proxy for pressure excesses associated with shocks and sites of particle acceleration associated with true radio hotspots. More studies taking into account particle acceleration are required to confirm whether these structures would genuinely form visible hotspots. 

The jets were visualised using {\tt yt}\footnote{\url{https://yt-project.org} } for Python. To do this we
converted the PLUTO computational volume, in spherical polars, to a
uniform three-dimensional $1025^3$ Cartesian grid using simple nearest-pixel
interpolation. The Cartesian grid is matched in resolution to the radial resolution of the spherical computational domain but does not resolve the details of the jet structures close to the injection region. It is however more than adequate to sample the spherical volume elements at large distances from the injection region. The Cartesian grids were read into {\tt yt} using the {\tt load\_uniform\_grid} function. 

A proxy for synchrotron emission in
the lobes was obtained by taking pressure to a fixed power (1.8) as
described in H20 and masking with a tracer threshold of $10^{-3}$. We used three distinct {\tt yt} functions: 

\begin{itemize}
  \item {\tt yt.create\_scene}: this is used to create
    three-dimensional contour maps of pressure in the lobes and the
    surrounding medium as seen from a point outside the computational volume. We use pressure rather than emissivity (where
    a tracer threshold is applied) for these maps because the
    discontinuities imposed by the tracer threshold cause problems for
    the three-dimensional contouring algorithm.
  \item {\tt yt.OffAxisSlicePlot}: this is used to create two-dimensional slices through the jet stream and two peaks in pressure within a 50 kiloparsec square. This slice was used to create plots of emissivity, pressure, density, velocity and divergence. 
  \item {\tt yt.ProjectionPlot}: three-dimensional line-of-sight projection onto a 2D plane of emissivity either along some arbitrary axis or constrained by the off-axis slices through the pressure maxima. 
\end{itemize}

\subsection{Radio observations}
The recent completion of LOFAR LoTSS DR2 \citep{shimwell22} generated a catalogue that covered 5,700 square degrees of the Northern sky and produced 4.4 million extragalactic radio sources. At the time of writing around 85 per cent of these sources have been associated with optical IDs through a combination of likelihood-ratio cross-matching and citizen science visual inspection through the Radio Galaxy Zoo (LOFAR) project. We used optically identified, large, bright sources from this catalogue to search for objects that exhibit signs of possibly precession-driven behaviour as discussed in Section \ref{sec:obs}.

\begin{table}
\caption{Table showing the events and representative simulations and timings. The simulation naming convention is precession cone opening angle, injection Mach number, precession period and resolution. }
\label{table:sim_summary}
\begin{center}
\begin{tabular}{ l r r r }
 Event & Simulation & Start frame & End frame \\
 \hline
 Hotspot Splitting& 45\_100\_1\_VHR & 42 & 66 \\ 
 Stream Splitting & 45\_100\_1\_VHR & 118 & 131 \\ 
 Chevron Spots & 45\_100\_02\_VHR & 78 & 98 \\ 
\end{tabular}
\end{center}
\end{table}

\section{Results}
\label{sec:results}

In the following subsections we describe in detail and discuss three of the many multiple hotspot events that are seen in our simulations. For each event we show three-dimensional contours of the pressure in the simulation as well as projected emissivity (the view seen by an observer from some arbitrary angle) and appropriate slices in physical quantities. The labelling of the simulations follows \cite{horton20b}. A given run will be denoted by $\psi$\_M\_pp\_res, where $\psi$ is the precession cone opening angle in degrees, M is the external Mach number, $pp$ is the precession period (1 meaning one complete turn per simulation time), and res denoting the resolution of the simulation where `VHR' stands for `very high resolution'. Whilst we ran simulations for a variety of settings (including jets with no precession), the ones presented in this paper are from more `extreme' precession parameters, such a wide angle for $\psi$ and / or rapid $pp$, as these show the new mechanisms more distinctly. In what follows we refer to a given set of simulation outputs (which are saved at regular simulation time intervals) as a `timestep'.

\subsection{Hotspot Splitting}
\label{subsec:sh}
\begin{figure*}
    \centering
    \includegraphics[width=0.9\linewidth]{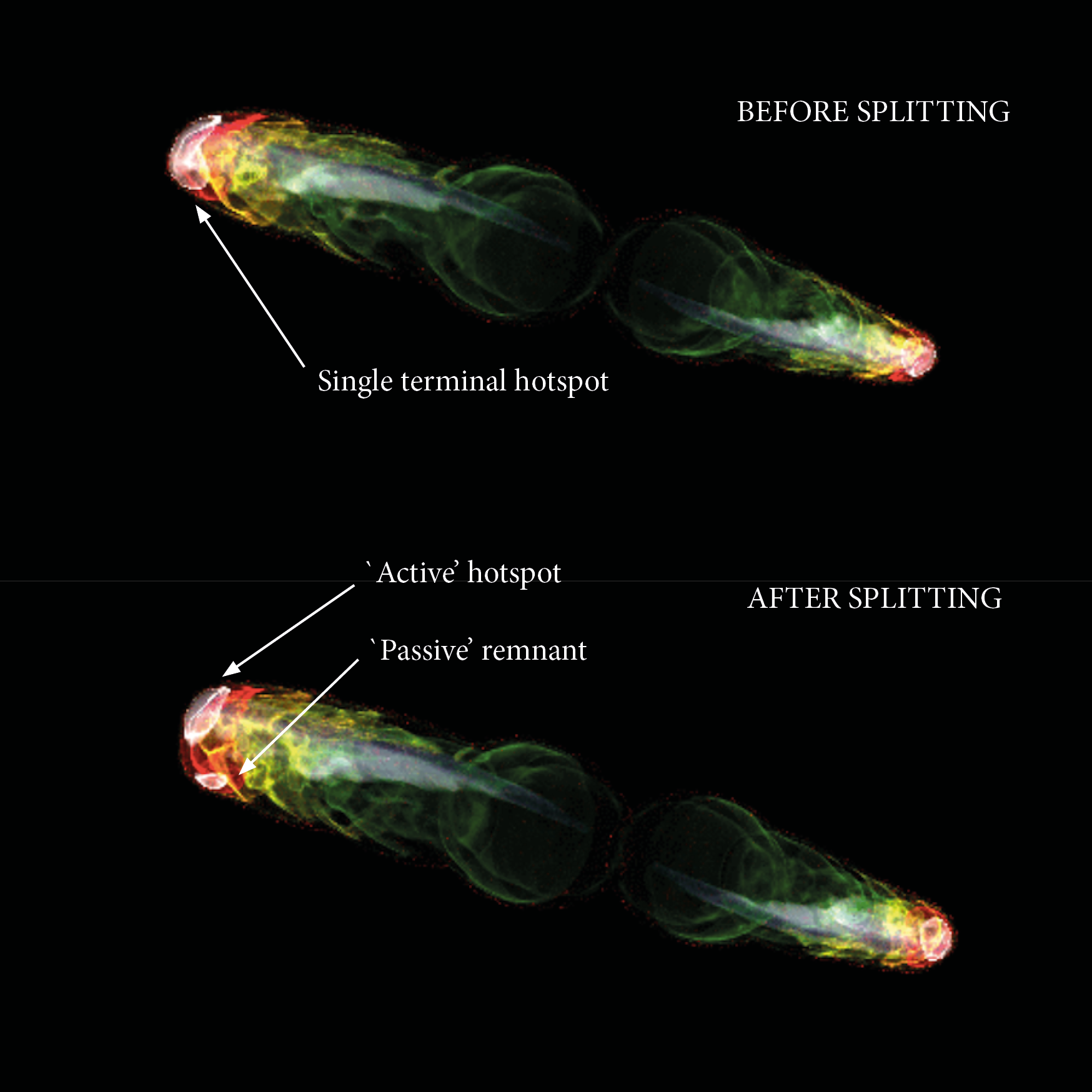}
    \caption{Top panel: a fast jet creates a FRII-style hotspot
      region where the jet terminates (timestep 44). Bottom panel: As
      the jet precesses, the hotspot breaks up (timestep 50). The
      smaller hotspot shrinks and recedes along the contact
      discontinuity as the lobe expands in the opposite direction. The pale shaded internal region represents the jet as determined from the Mach number. Isobaric surfaces are used to visualise lobes and hotspots. The white contours at the edge of the lobes are the regions of highest pressure and are used to indicate hotspots.}
    \label{fig:double_contour}
\end{figure*}

\begin{figure*}
    \centering
    \includegraphics[width=0.9\linewidth]{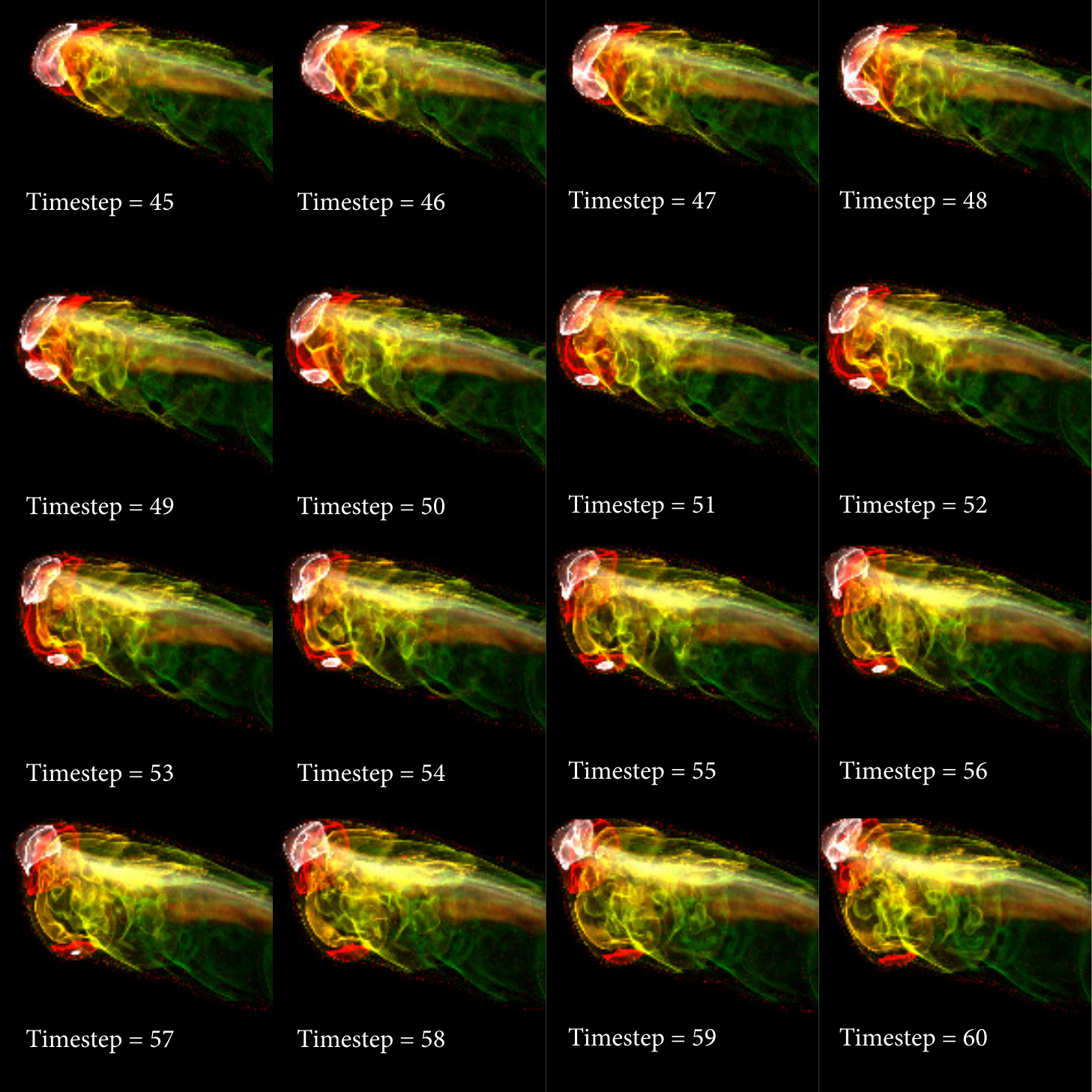}
    \caption{As Fig~\ref{fig:double_contour}, showing the evolution of the hotspot splitting event, running from timesteps 45 to 60 representing a duration of $1.8 \times 10^6$ years.}
    \label{fig:hotspot_evolution}
\end{figure*}

\begin{figure*}
    \includegraphics[width=0.95\linewidth]{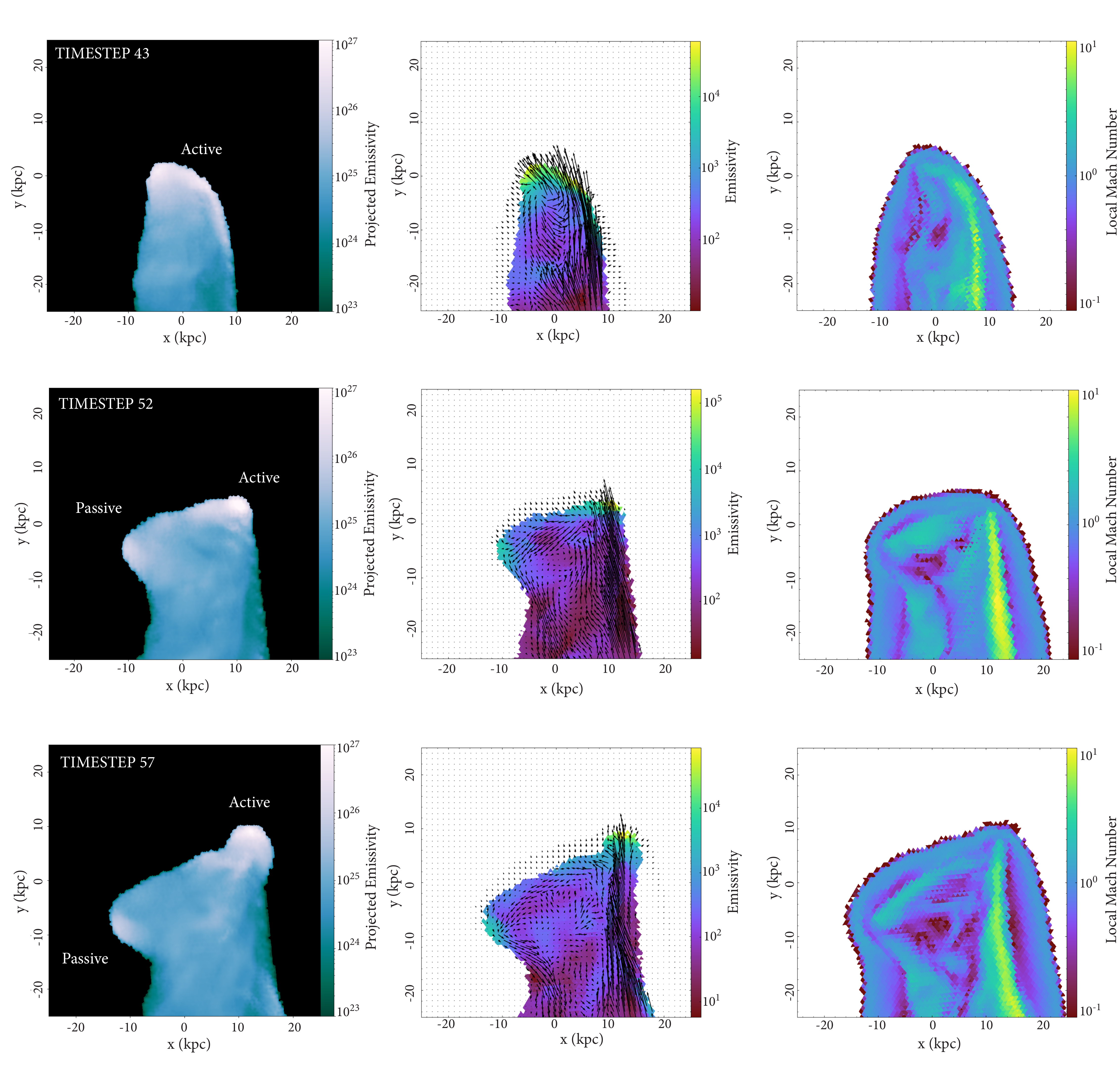}
\caption{Evolution of hotspot development from beginning of hotspot separation event (top row, simulation timestep 43), middle (middle row, simulation timestep 52) to end (bottom row, timestep 57). First column shows projected emissivity. Second column shows emissivity slices with overlaid velocity vectors. Third column shows the local Mach number. The slices correspond to the south lobe seen in the pressure contours of Fig.~\ref{fig:double_contour}. The contact boundary has been masked out of the emissivity to show only processes internal to the lobe.}
\label{fig:double_slices}
\end{figure*}

For this event, in simulation 45\_100\_1\_VHR, during the time range specified in Table \ref{table:sim_summary}, the simulation initially shows a large terminal hotspot forming as would be expected for an FRII radio jet. As the jet precesses and the lobe expands, this hotspot splits into two components. We name these the `active' and `passive' hotspots, $h_A$ and $h_P$, and they can be seen in Fig.\ \ref{fig:double_contour}, which shows pressure contours from the beginning and end of the hotspot splitting event. The full evolution of this event can be seen in \ref{fig:hotspot_evolution}.
Fig.~\ref{fig:double_slices} shows a $3\times 3$ grid of the
development of the hotspot splitting event from the first timestep
where two distinct peaks can be found (top row, time 43), through to a point
close to the end where $h_P$ is noticeably diminished but still
visible (time 57); here the first column shows modelled emissivity
of the whole source, the second column shows a slice in emissivity
through the two hotspots with overplotted velocity vectors, and the
third shows the same slice through the divergence of the velocity field. As can be
seen in these images, the `active'
hotspot remains at the jet terminus and is the site of the continued
jet termination shock. The
passive hotspot $h_P$ is not an ongoing shock or significant site of
compression (negative divergence) but is simply a remnant of the
original large overpressured hotspot. The outflow from $h_A$ pushes $h_P$
along the inside of the contact discontinuity back down towards the
core while continuing to confine it. Over time the central compact region of the
passive hotspot $h_P$ shrinks, whilst the rest of it expands and
diminishes until $h_P$ is no longer visible. 

\begin{figure*}
\includegraphics[width=0.8\linewidth]{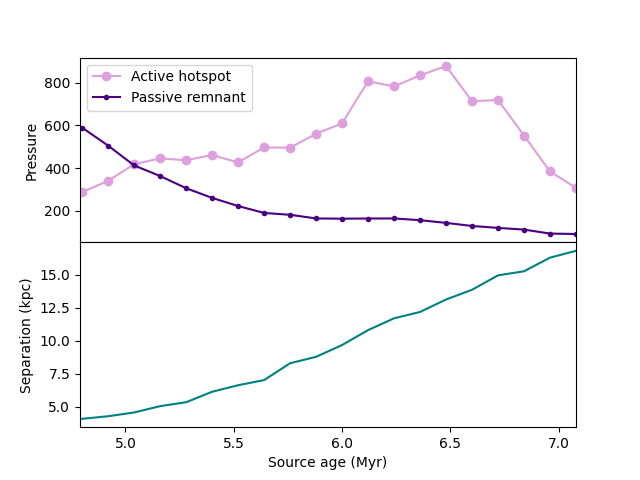}
\caption{Evolution in active ($h_A$) and passive ($h_P$) hotspots before and after hotspot splitting. Top panel shows evolution of pressure in simulation units in the `active' hotspot (pink) and `passive' remnant (purple). Bottom panel shows increasing distance between hotspots as the splitting event continues.}
\label{fig:double_pressure}
\end{figure*}

The evolution of the peak pressure in the two hotspots and their
separation is shown in Fig.~\ref{fig:double_pressure}, where we see the pressure in the passive hotspot falling to equal the ambient pressure at the head of the lobe over the duration of the event. Following
\cite{cox91} we define the adiabatic expansion timescale as the sound
crossing time for the secondary hotspot, $t=d/v_s$ where $v_s$ is the
local sound speed. In physical units as defined in our earlier paper \citep{horton20b} the hotspot
diameter $d$ is around 1.5 kpc and the local sound speed is $\sim 30$
times the simulation unit of speed, or $1.4 \times 10^{4}$ km
s$^{-1}$. This gives an adiabatic expansion timescale of $\sim 10^5$
years, which is comparable to a single timestep of the simulation for a hotspot of this size.
Clearly as the passive hotspot persists for $\sim 2 \times 10^6$
years, as shown in Fig.~\ref{fig:double_pressure}, it is not freely
expanding, consistent with the idea that the outflow from $h_A$
confines $h_P$ without generating an active shock. Contrary to the models discussed in Section \ref{sec:intro} we see that the compact hotspot neither signifies the previous location of the jet nor is it formed by a shock in the backflow. This `hotspot splitting' mechanism is related to, but distinct from, to the `dentist's drill' and `splatter spot' hotspot formation models. For example, there is another disconnection event starting at Timestep 59, which could be described in terms of the dentist's drill model since a new jet termination point is established and the jet no longer reaches the site of the previous hotspot. In this event, the adiabatic expansion results in a fading time of around five timesteps, which is expected given the much larger size of the remnant; this is much shorter than the persistence time of the passive hotspot in the hotspot-splitting scenario.

We note that the hotspot splitting process described here does not have to be driven by
jet precession (we have observed it on very short periods during early stages of non-precessing simulations), but it is more noticeable and longer-lasting in slowly precessing jets
(our $\_1$ simulations), and has a distinct evolution that is absent in non-precessing sources. The wider the precession cone opening angle
$\psi$, the more noticeable the hotspot splitting and the more
pronounced the differences in physical properties between $h_A$ and $h_P$. Therefore the greater the differences between the two hotspots, the more
plausibly they are associated are with a strongly precessing source.
Many double hotspots seen in well-known radio sources (e.g. 3C\,173.1 N,
which strikingly resembles the emissivity plots of Fig.~\ref{fig:double_pressure},
or 3C\,20 E, \citep{hardcastle98}) could be explained at least as well by
this mechanism as by previously described models such as splatter-spot or
dentist's drill.

\subsection{Stream Splitting and Velocity Structure}
\label{subsec:ss}
\begin{figure*}
    \centering
    \includegraphics[width=0.9\linewidth]{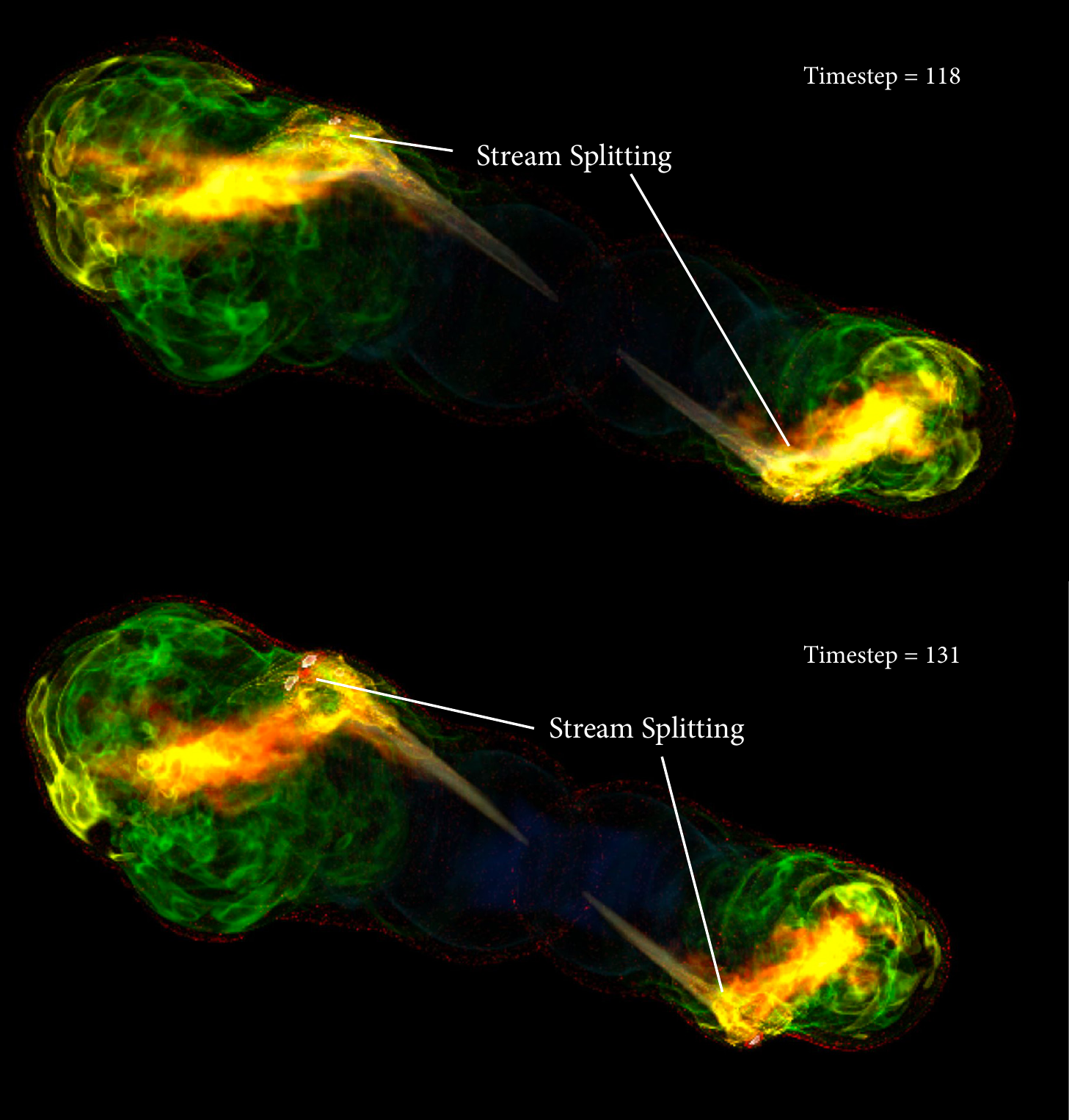}
    \caption{As in Fig.~\ref{fig:double_contour} but showing a `stream splitting' event later in the same simulation (timestep 118) Bottom panel: as above,  but from timestep 131. Here the Mach number of the flow is represented by the yellow colours and the regions of highest pressure (hotspots) are represented by the smaller white contours. Some hotspots are hidden by projection and backflow. 
    }
    \label{fig:splatter_contour}
\end{figure*}

The second event of Table \ref{table:sim_summary} shows one of many
regular `stream splitting' events, illustrated in the pressure
contours of Fig.~\ref{fig:splatter_contour}. During this time period the jet interacts with the
 lobe boundary, resulting in the entrainment of dense material which breaks the jet path into two or more component parts which then continue travelling almost in parallel inside the lobe. 

\begin{figure*}
  \includegraphics[width=0.95\linewidth]{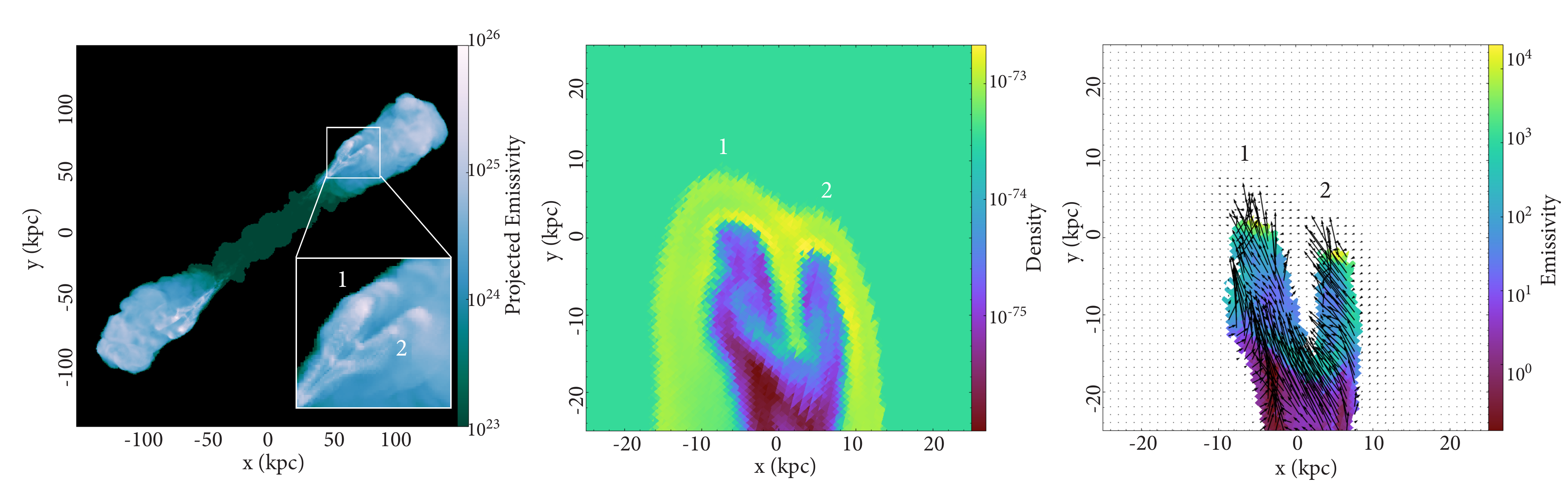}

\caption{Caption as in Fig.~\ref{fig:double_slices}, showing the stream splitting event from simulation 45\_100\_1\_VHR at timestep 118. Columns represent projected emissivity (first column) including a cutout of the approximate location of the slices, density slice (middle column) and emissivity slice with velocity vectors (last column). This is one snapshot of an event which corresponds to a duration in physical units of approximately $1 \times 10^7$ years and is just one of many stream-splitting events in this simulation.}
  \label{fig:splatter_slices}
\end{figure*}

 Fig.~\ref{fig:splatter_slices} shows the formation of two terminal
 hotspots from this split stream, with the columns being (left) the
 projected emissivity of the whole source, and (middle) density and
 (right) emissivity slices through the double hotspot in the lower
 half of the image. Whilst one  hotspot is slightly larger than the other, as being connected to more  of the high-velocity flow, both are still active termination shocks both in emissivity and in the divergence of the velocity field (not shown). The compression remains throughout the event, for as long as the jet remains split. The splitting is caused by an overdense region of the shocked external medium forcing the jet apart (see density plot in Fig.\ \ref{fig:splatter_slices}), but
 this structure can be persistent and long-lasting. The hotspots show little morphological evolution throughout the event, which lasted for approximately $\sim 2 \times 10^6$ years. Additionally, the jet does not have to be split into only two components: during one `slowly' precessing simulation, the jet split into three distinct streams which broke apart and recombined for around one third of the total simulation time. Although in this particular case the hotspots appear low down in the lobe, this is the result of projection combined with the precessing jet: in this simulation the jet is now pushing through previously undisturbed material and so the hotspots are effectively at the end of a newly-emerging lobe. The interaction with the shocked external medium may be mediated by fluid dynamical instabilities at the end of the jet; our simulations do not allow us to identify the exact mechanisms (e.g. Rayleigh-Taylor or Kelvin-Helmholtz instability at the contact discontinuity) but this issue would merit further study in simulations making use of realistic environments.

In the projected emissivity maps (Fig.~\ref{fig:splatter_slices}, col. 1)
we see that the two hotspots can be difficult to
distinguish from the surrounding jet material; in this case they appear as two
bright filaments extending out from the jet. These might be
interpreted as hotspots, jet knots, or lobe filaments in a real-world
radio map: of course, without modelling particle acceleration it is
difficult to tell how these would realistically appear, but we can see from
our plots that they produce a pressure excess and are the sites of active shocks. 

As a consequence of the boundary interaction that gives rise to
the double hotspots, we also observe complex internal velocity
structure. 
In the later stages of simulations of jets with wide
precession cone angles such as this one, we find persistent structures -- Mach `clouds' -- where the velocity remains high even long after the jet has precessed away. This is due to the turbulent backflow being deflected by the jet's initial contact with the lobe boundary right into the centre of the lobe. This is shown by the yellow Mach structures visible in Fig.~\ref{fig:splatter_contour}. These occur whether or not hotspots have formed so long as the flow is fast enough. It is important to note that these flows produce something akin to `splatter spots' as they result in secondary shocks forming on the opposite side of the lobe. The fact that twin hotspots can create downstream splatter spots highlights that many of these features may be interrelated. The outflow from the twin active hotspots is poorly collimated, resulting in a less distinct, and more diffuse, splatter spot. 

The key distinguishing feature of stream splitting would be that
particle acceleration would be equally active in two hotspots of
similar appearance (unlike the
case in hotspot splitting, above, or in dentist's drill or
splatter-spot models) and this could explain the detection
of X-ray synchroton emission, an indicator of ongoing particle
acceleration, in several well-studied double hotspots, such as in 3C\,227
\citep{hardcastle07}. The splitting in the jet itself would only be
visible if emissivity and Doppler boosting favoured it, but there are
some objects that could be interpreted in this way, most recently in
the images of CGCG 021$-$063 by \cite{fanaroff21}.

\subsection{Chevron Spots}
\label{subsec:chevron_spots}
\begin{figure*}
    \centering
    \includegraphics[width=0.8\linewidth]{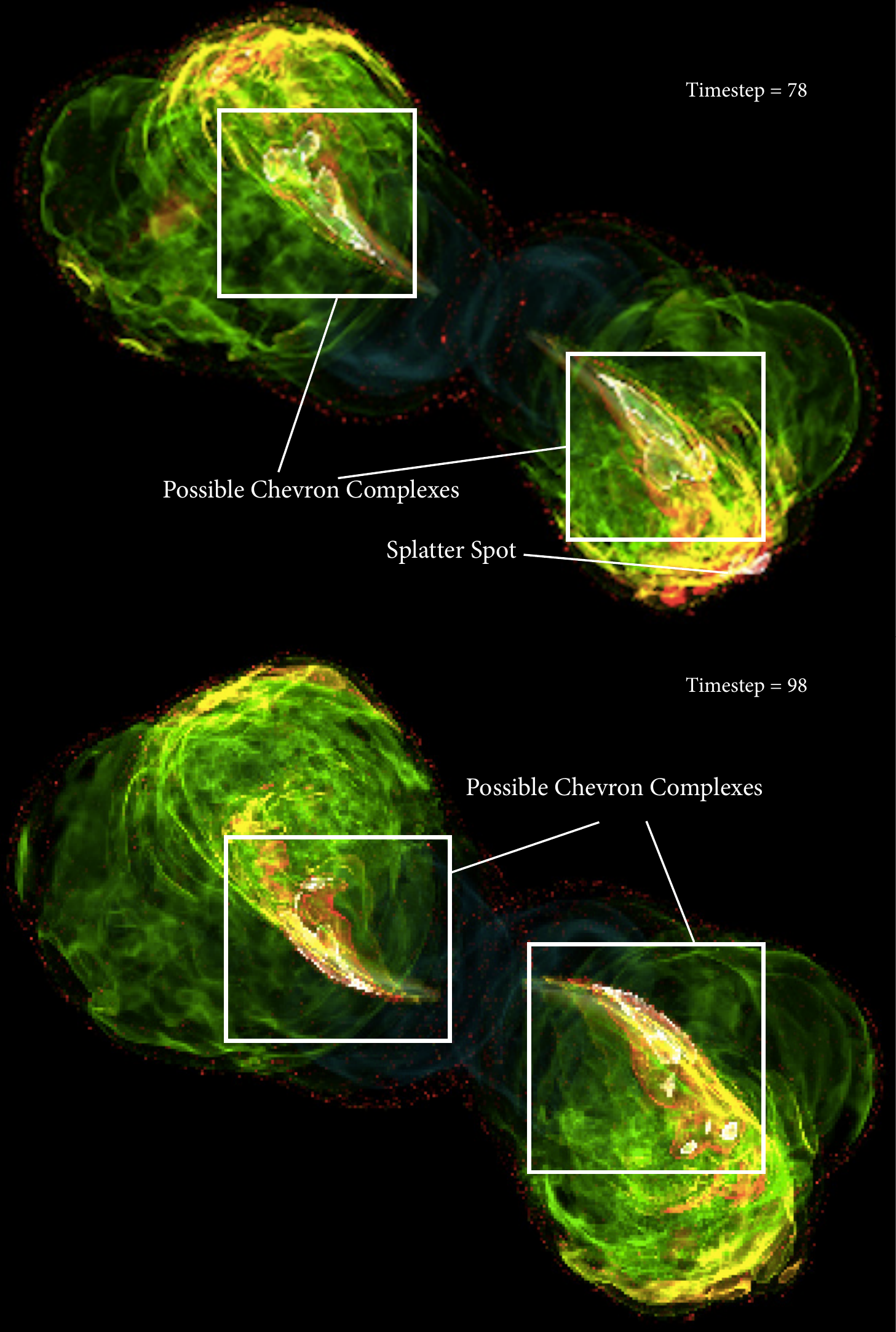} 
    \caption{Top panel: white contours show areas of shocks forming chevron-shaped hotspots moving away from the core (simulation 45\_100\_02\_VHR, timestep 78), with possible splatter spot shown at the edge of the lobe. Bottom panel: as above, but the hotspot begins to break up as the precessing jet moves rapidly away from its previous location (timestep 98). 
    }
    \label{fig:chevron_contour}
\end{figure*}

\begin{figure*}
    \includegraphics[width=0.90\linewidth]{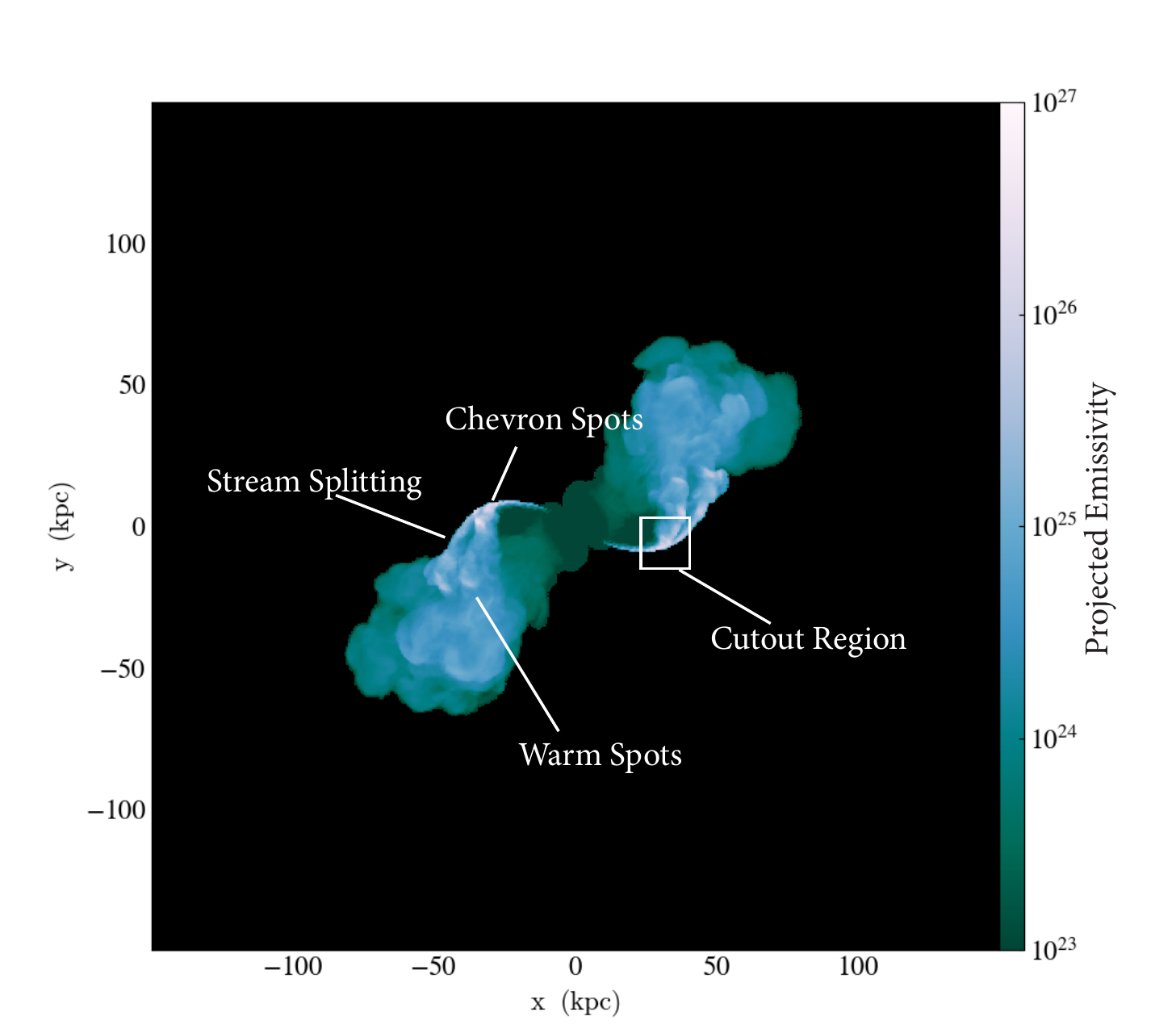}
  \includegraphics[width=0.95\linewidth]{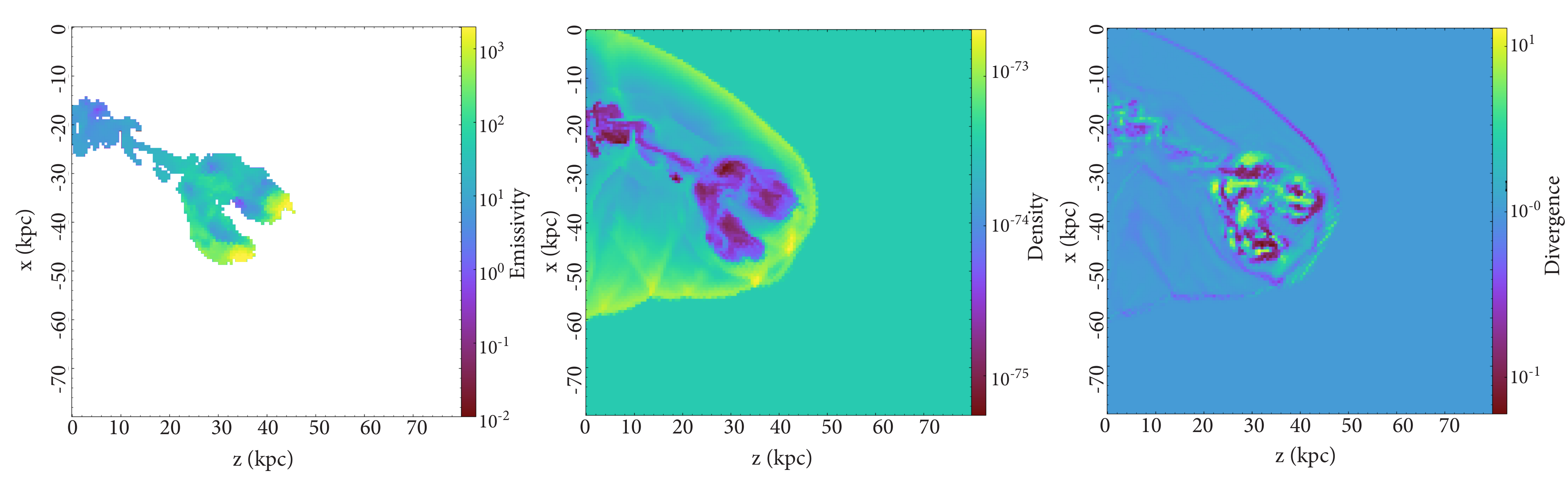}
  
\caption{Projected emissivity of chevron structure showing large hotspot complexes in both lobes that would likely be interpreted as bright curved jets. (top panel). Slices through emissivity maximum (bottom left), plus density (bottom middle) and divergence (bottom right) at the same slice position as for the emissivity.}
  \label{fig:chevron_divergence}
\end{figure*}

In the final example that we select, from the simulation
45\_100\_02\_VHR (Table \ref{table:sim_summary}) the `extreme'
precession of a rapidly precessing jet with a wide precession cone
opening angle creates a highly complex and unstable environment.
Striking chevron-shaped hotspot regions -- which are more akin to dynamic systems or complexes where multiple short-lived hotspots arise from a larger dynamical structure -- appear at the base of disrupted
jets/plumes, as seen in Fig.\ \ref{fig:chevron_contour}, which shows
the pressure contours for the whole source, and the top panel of
Fig.\ \ref{fig:chevron_divergence}, where the projected emissivity is shown. Like the stream splitting events of the previous subsection, the chevrons appear to be created as the precession drives an already-splitting jet into the edge of the lobe. However, these features are distinct because, rather than creating one or more terminal hotspots, the jet transitions to a series of shocks, as can be seen in emissivity, density and divergence slices through the curved shock regions (Fig.~\ref{fig:chevron_divergence}, bottom panel). Further into the lobes, the observed structure is unstable and multiple short-lived `hotspots' are visible in emissivity in some timesteps
(Fig.\ \ref{fig:chevron_contour}, bottom panel).

The projected emissivity maps do not consistently preserve the underlying
structure that appears in the pressure contours but instead produce highly curved, bright jets that break up often and appear knotty.
These often terminate in two distinct hotspot regions. These would be difficult to identify in radio maps resulting in such complex systems going unnoticed. Moreover the curvature
of the jet and its brightness are dependent upon the viewing angle of
the projected source, rather than any feature of the simulations
themselves: the same jet can appear straight from a different angle,
and the chevron structures themselves are highly dynamic and unstable.
Whilst `v-shaped' or chevron-shaped hotspots are not commonly observed
in extragalactic sources, they would only be distinguishable
from other structures at high resolution, and there are many objects with
bright, curved, and / or discontinuous jets terminating in multiple
hotspots. Therefore it is possible that such features may be overlooked as an indicator of more extreme precession. At low resolution, we might see
the chevrons as bright `hotspots' at the base of the lobes combined
with complex lobe structure further out, and examples of this
morphology are not uncommon in powerful radio sources, e.g. 3C\,215 or 3C\,249.1
\citep{bridle94}. The large-scale structure of IRAS J1328+2752
\citep{nandi21}, a known binary black-hole system, is another example
of this class.

\subsection{Prevalence of Different Mechanisms}
\label{subsec:prevalence}

\begin{table}
\caption{Table showing the percentage of time that any of the five stated mechanisms for multiple hotspot formation are visible, plus a category for complex events where it is not possible to identify a formation mechanism. These percentages are taken by visual inspection from a single fixed view and may under-represent the true time multiple hotspots are present. *See Section \ref{subsec:prevalence} for discussion.}
\label{table:multi-hotspots}
\begin{center}
\begin{tabular}{ l r r r }
 Mechanism          & \_STR                 & \_1             & \_02\ \\
 \hline
Unknown             &  3.75\%               & 5.2\%             & 8.8\% \\
Hotspot Splitting   &  5.6\%                & 4.9\%             & 5.10\% \\
Stream Splitting    &  27.5\%               & 34.6\%            & 23.8\% \\
Chevron Spots       &  0                    & 0                 & 58.1\% \\
Splatter Spots      &  0                    & 0                 & 4.8\%\\
Dentist's Drill     &  0                    & 3.8\%             & 0 \\
\hline

Frames with Multi Hotspots &  52                   & 138             & 262 \\
Frames in Simulation       &   160                 & 286             & 294 \\
\hline 
 Time Hotspot Visible            & 32.5\% *                & 48.2\%          & 89.1\% \\
Time Visible (years)         &  $6.2 \times 10^6$  &  $1.7 \times 10^7$  & $3.1 \times 10^7$     \\
Source Age (years)           &  $1.9 \times 10^7$  &  $3.4 \times 10^7$  & $3.5 \times 10^7$     \\
\end{tabular}
\end{center}
\end{table}

\begin{figure*}
\includegraphics[width=0.8\linewidth]{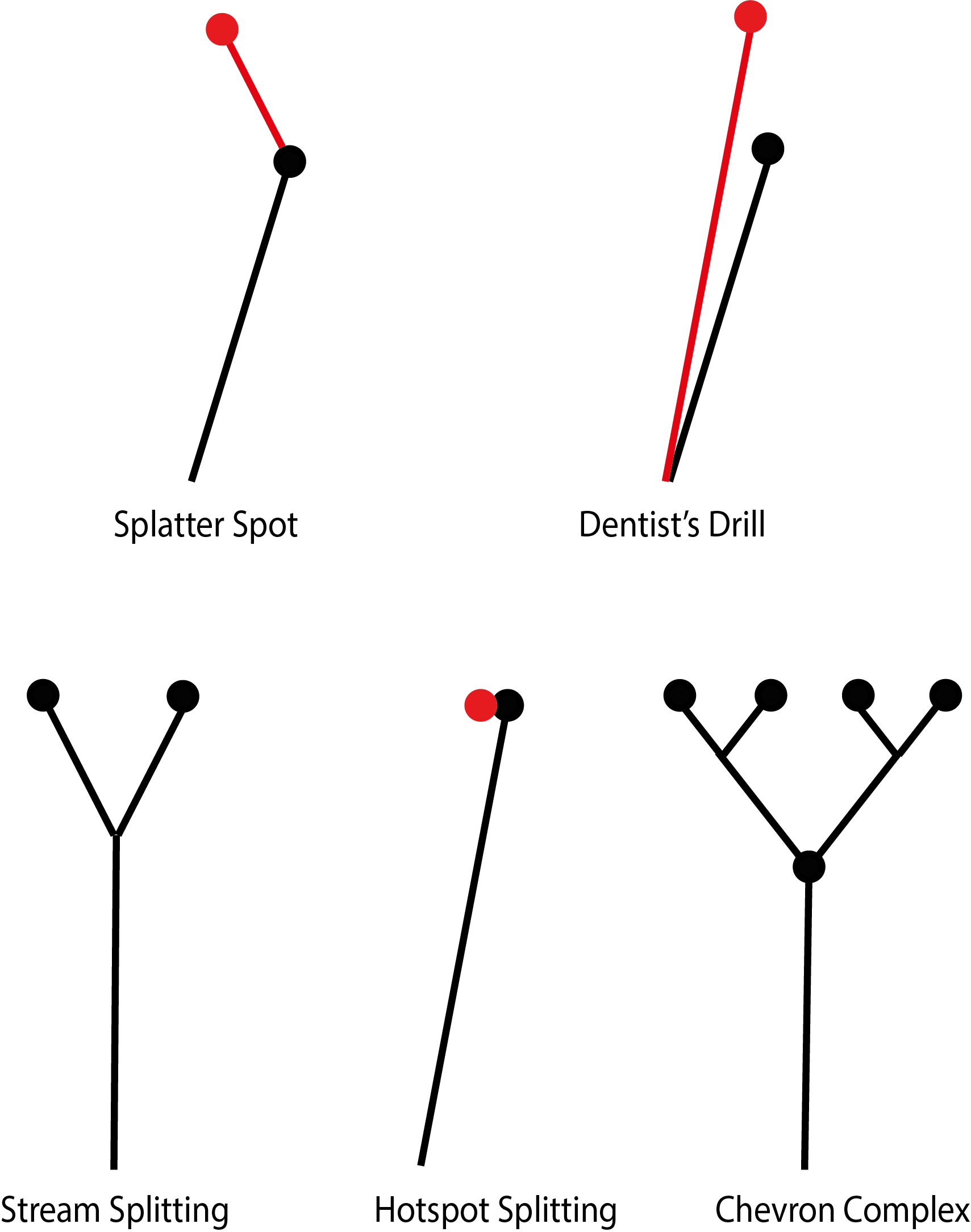}
\caption{Schematics showing the development of five mechanisms for multiple hotspot generation. Lines indicate the jet and circles indicate the location of terminal hotspots. The lines in black are caused by the current jet flow and red indicates flow and hotspots distinct from the current jet. TOP ROW: two mechanisms for multiple hotspot formation described in the current literature: splatter spot (left), where a primary flow (black) hits the lobe boundary and deflects into a secondary flow (red); dentist's drill (right), where the current flow (black) is distinct from the previous flow (red). BOTTOM ROW: the three mechanisms described in this paper; stream splitting (middle) where the jet splits into a stable and long-lasting double flow; hotspot splitting (middle) where a large terminal hotspot breaks in two and the remnant hotspot (red) is slowly forced apart by hydrodynamic forces; chevron complexes (right) where the system generates short-lived, chaotic shocks from an unstable flow structure.} 
\label{fig:jet_diagram}
\end{figure*}

Figure \ref{fig:jet_diagram} shows the schematic of five mechanisms for the creation of multiple hotspots: the existing `Dentist's Drill' and `Splatter Spots' plus the three detailed in this paper. It is important to note that these mechanisms are not mutually exclusive. Precessing jet hydrodynamics cause complex internal lobe structures; we observe that more `extreme' simulations (those with shorter precession periods and thus multiple turns throughout the source age) show overlapping events where clusters of multiple hotspots occur in different parts of the lobe from apparently different mechanisms (such as chevron spots leading to splatter spots, and so on).

Table \ref{table:multi-hotspots} shows the prevalence of different formation mechanisms of multiple hotspots in the three different simulations. We have included an `unknown' category where it difficult to tell where the hotspots are coming from, particularly early in the simulation, or where there is ambiguity in the flow structure. The result of this table appear to show a misleadingly high prevalence of multiple hotspot formation in the non-precessing simulation -- however, this occurs very early on in the simulation before the jet becomes stable and produces a single terminal hotspot. The straight jets also expand quickly to run off the grid which means the simulation only runs for half of the length of the other two simulations, so the 32.5\% prevalence of multiple hotspots comes from a much shorter simulation and is plausibly an overestimate by a factor $\sim 2$. Since the non-precessing jet only shows large, close multiple hotspots early in its lifetime, this is likely caused by interactions with the environment on scales smaller than the jet collimation length scale, but more simulations are required to confirm this.

The duration and complexity of multiple hotspot generation increases with precession period. No other parameters are varied. The rapidly precessing ($\_02$) simulations show multiple hotspots for almost 90\% of the simulation time, with jets moving from, for example, hotspot splitting to stream splitting as the jet precesses and breaks through the lobe boundary. Rapid precession also brings an increase in unclassifiable events, which are often too small and short-lived to analyse properly. We observe clusters of hotspot formation in the rapidly precessing clusters that have more than one cause: hotspot complexes close to the jet likely have a different mechanism for their formation than, say, those at the end of the lobe even when occurring in the same frame. We also apparently see a weak evolution of these mechanisms over time, with hotspot splitting generally occurring at earlier times and long periods of stable stream splitting occurring in the middle of the source lifetime. `Chevron' spots only occur in the rapidly precessing simulation, but can occur at any point in the lifetime of the source.

It is worth mentioning that the mechanisms listed in \ref{table:multi-hotspots} are only recorded if they generate strongly overpressured regions. There are long-lasting events of jet deflection associated with the formation of splatter spots. These do not form `hotspot' structures in our simulations but observationally may be visible and associated with `warm spots' (e.g., \citep{leahy97}).

\section{Comparison to observations}
\label{sec:obs}

Precessing jets are a useful tool in the hunt for binary supermassive black hole systems and produce predictable morphological changes to jet paths. In recent years, these have been used to create estimates of the binary separation and gravitational wave strain of certain sources \citep{krause18,horton20a,horton20b}. \cite{horton20a} in particular highlighted the importance of accurately identifying terminal hotspots in determining a jet path that can be used to constrain binary black hole separation. The mechanisms described above were common in our high-resolution simulations, suggesting they could be common features of precessing jets. 

As a proof of concept, we used the second data release of the LOFAR Two-metre Sky Survey (LoTSS) \citep{shimwell22} with preliminary optical identifications to search 10,000 extended sources for a population of large ($>2$ arcmin), bright ($> 50$ mJy) sources with optical identifications showing precession indicators, namely jet curvature, S-symmetry, jet-lobe misalignment and multiple hotspots, see \cite{horton20b} for details. Due to resolution issues we focused on just two of these indicators, which were curvature and multiple hotspots. After visual inspection of  a sample of more than 2,000 candidate sources we selected 112 which showed the above two precession indicators. This population shows a high prevalence of features which can be observed in our simulations.

\begin{figure*}
    \includegraphics[width=0.90\linewidth]{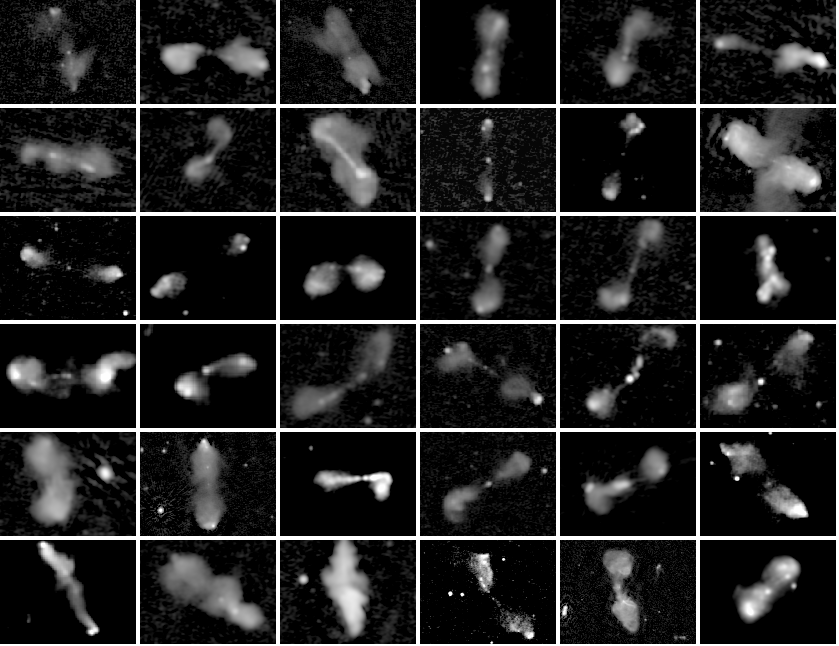}
\caption{Selection of precessing jet candidate sources in LOFAR LoTSS DR2. Greyscales show total intensity at 144 MHz with a resolution of 6 arcsec. The sources show a selection of precession indicators such as curved jets and multiple hotspots. Numbering used in the text uses rows first followed by columns, with the top row and first column on the left being labelled (1,1), and the last image in the bottom row being (6,6).}
  \label{fig:lofar_lotss}
\end{figure*}

Fig.\ \ref{fig:lofar_lotss} shows a selection of 36 radio galaxies taken from the sample of 112 described above. Most of these show some suggestion of multiple hotspots or warm spots whether or not they have observable continuous jets. We have not tried to identify the specific mechanisms responsible for multiple hotspot formation, as this would necessarily be subjective and inaccurate. However, some hotspot clusters map more readily onto some of the mechanisms included in the paper; for example, the images shown in (4,2), (2,6) and (6,6) may be indicative of hotspot splitting, and other mechanisms may be present in different images. A detailed exploration of precession characteristics for this population will be covered in a subsequent paper, in which we will make use of higher-resolution radio images and the spectra of hotspots in order to suggest formation mechanisms for individual cases.

The availability of the LOFAR data opens up the possibility in future of testing the suggestions of \cite{krause18} and \cite{horton20a} with a much larger database. By exploiting the very large databases of resolved radio sources to be provided by future large-area radio surveys, we may be able to constrain the merger history of supermassive black holes, a key prediction of detailed models of galaxy evolution.

Jet precession may also have an important role to play in the feedback effects of radio AGN, which are now widely thought to influence the evolution of the galaxy mass function by suppressing cooling in their hot haloes \citep{mcnamara12}. In our simulations we see that the lobes of precessing jets grow more slowly than non-precessing ones, as jet momentum flux is spread over a larger area \citep[e.g.,][]{horton20b}. This means that for a given source age, precessing jets heat material closer to the galactic centre. This may solve a long-standing problem whereby non-precessing jets in simulations tend to spend most of their time heating material with low cooling rates far from the galaxy centre \citep[e.g.,][]{omma04,hardcastle13}. If so, again, it is crucial to identify sources in which precession is taking place using the indicators highlighted in this paper and our previous work.

The processes described in this paper come from simulations designed to mimic plausible precession periods from supermassive black hole binaries. However, there are other ways in which these disturbances can occur on both short and long timescales. Accretion disks can influence black hole spin direction through the Bardeen-Petterson effect \citep{bardeen75,lense18}, which results in the inner accretion disk aligning with black hole spin axis. Given the chaotic environments at the centres of AGN-producing galaxies, jets can experience perturbations without the presence of a supermassive binary \citep[e.g.,][]{liska19}. Accretion disks are chaotic environments; whether Bardeen-Petterson precession might dominate earlier in a source lifetime, and thus be more related to features seen in smaller or younger jets, remains unknown.

\section{Conclusion}
\label{sec:conclusion}

Using high-resolution numerical simulations of precessing jets, we
have identified three new mechanisms that can produce complex multiple
hotspots in powerful radio galaxies. It is important to note that
these three mechanisms should be seen as being
additional to, rather than replacing, current theories of multiple
hotspot formation such as splatter spots and the dentist's drill
model. Both of these prior mechanisms can be seen in our simulations
alongside hotspot splitting, stream splitting and chevron structures.
Whilst all three novel mechanisms are linked to precession,
particularly in the changes in lobe structure that arise as a
consequence, some of these mechanisms -- e.g., hotspot splitting --
may also occur in non-precessing sources, but to a less extreme degree
and for shorter duration. Our work thus supports the argument of \cite{krause18} that multiple hotspots in general can be indicators of jet precession, for example as a result of binary supermassive black hole activity. 

This paper focuses on the consequences of `extreme' precession,
specifically a 45$^\circ$ precession cone opening angle. However the
listed mechanisms -- perhaps with the exception of chevron structures
-- can also be found in sources with less extreme parameters. In such
cases they may resemble real-world sources (such as Cygnus A, Hercules
A and Hydra A) with multiple hotspots and more typical lobe structure.
Given the prevalence of multiple hotspots in real-world sources -- as seen by the increase in suitable candidates in LOFAR LoTSS DR2 -- more work is necessary to understand which of these possible mechanisms, if any, are at work. A more robust understanding will lead to more realistic constraints (such as binary black hole properties, merger history and feedback processes) on the nature of the radio-loud AGN population. 

\section*{Acknowledgements}

MAH acknowledges a studentship from STFC [ST/R504786/1] and support from STFC grants [ST/R000905/1] and  [ST/X002543/1]. MJH acknowledges support from STFC  grants [ST/R000905/1], [ST/V000624/1] and [ST/X002543/1]. Simulations were performed on the University of Hertfordshire High Performance Computing cluster. \footnote{\url{https://uhhpc.herts.ac.uk/}.}  Radio observations were taken from the LOFAR Two-metre Sky Survey Data Release 2 (LOFAR LoTSS DR2). LOFAR (the Low Frequency Array) was designed and constructed by ASTRON and operated by the ILT foundation under a joint scientific policy. The ILT resources have benefited from the following recent major funding sources: CNRS-INSU, Observatoire de Paris and Université d'Orléans, France; BMBF, MIWF-NRW, MPG, Germany; Science Foundation Ireland (SFI), Department of Business, Enterprise and Innovation (DBEI), Ireland; NWO, The Netherlands; The Science and Technology Facilities Council, UK; Ministry of Science and Higher Education, Poland; The Istituto Nazionale di Astrofisica (INAF), Italy. The authors would like to express their sincere gratitude to the anonymous reviewer for their very helpful suggestions for improving this paper.

%%%%%%%%%%%%%%%%%%%%%%%%%%%%%%%%%%%%%%%%%%%%%%%%%%
%%%%%%%%%%%%%%%%%%%% REFERENCES %%%%%%%%%%%%%%%%%%
%%%%%%%%%%%%%%%%%%%%%%%%%%%%%%%%%%%%%%%%%%%%%%%%%%

\section*{Data availability}
No new observational data were generated or analysed in support of this research. Simulation source files are available on request. LoTSS DR2 images can be downloaded from \url{https://lofar-surveys.org/}.

\bibliographystyle{mnras}
\bibliography{main} % if your bibtex file is called example.bib

\bsp	% typesetting comment
\label{lastpage}
\end{document}